\documentclass[DIV12,11pt]{scrartcl}

\usepackage{amsthm,amssymb,amsmath}
\usepackage[noblocks]{authblk}

\setkomafont{title}{\normalfont \LARGE \bfseries}
\setkomafont{section}{\normalfont \Large \bfseries \boldmath}
\setkomafont{subsection}{\normalfont \large \bfseries}
\setkomafont{subsubsection}{\normalfont \normalsize \bfseries}
\setkomafont{descriptionlabel}{\itshape}
\setkomafont{paragraph}{\normalfont \bfseries \boldmath}

\newtheorem{theorem}{Theorem}
\newtheorem{lemma}[theorem]{Lemma}
\newtheorem{corollary}[theorem]{Corollary}

\newtheorem{definition}[theorem]{Definition}

\usepackage{hyperref}
\usepackage{hyperref}
\usepackage{amsmath,amsthm}
\usepackage{stmaryrd}
\usepackage{relsize}

\usepackage{amssymb}
\usepackage{graphicx}
\usepackage{color}
\usepackage{algpseudocode}
\newcommand{\remove}[1] {}

\newlength{\pgmtab}
\newenvironment{program}{%
\begin{tabbing}\hspace{0em}\=\hspace{0em}\=%
\hspace{\pgmtab}\=\hspace{\pgmtab}\=\hspace{\pgmtab}\=\hspace{\pgmtab}\=%
\hspace{\pgmtab}\=\hspace{\pgmtab}\=\hspace{\pgmtab}\=\hspace{\pgmtab}\=%
\+\+\kill}{\end{tabbing}}

\title{An alternative proof for the constructive Asymmetric Lov\'asz Local Lemma\thanks{Research co-financed by the European Union (European Social Fund ESF) and Greek national funds through the Operational Program ``Education and Lifelong Learning'' of the National Strategic Reference Framework (NSRF) - Research Funding Program: ARISTEIA II.}}

\author[1,2]{Ioannis Giotis}
\author[1,2]{Lefteris Kirousis}
\author[1]{Kostas I. Psaromiligkos}
\author[1,2,3]{Dimitrios M. Thilikos}

\affil[1]{Department of Mathematics,  National Kapodistrian University of Athens, Greece 
and
}
\affil[2]{Computer Technology Institute Press  ``Diophantus'', Patras, Greece}
\affil[3]{AlGCo project-team, CNRS, LIRMM, France}

\date{\vspace*{-2em}}

\begin{document}

\maketitle

\thispagestyle{empty}

\vspace{-10mm}
\begin{abstract}
\noindent We provide an alternative constructive  proof of the Asymmetric Lov\'asz Local Lemma.
Our proof uses the classic algorithmic framework of Moser and the analysis introduced by Giotis, Kirousis, Psaromiligkos, and Thilikos in ``On the algorithmic Lov\'asz Local Lemma and acyclic edge coloring", combined with the work of Bender and Richmond on the multivariable Lagrange Inversion formula. 
\end{abstract}

\vspace{-7mm}
\section{Introduction}
\vspace{-3mm}

The Lov\'{a}sz Local Lemma (LLL) is a powerful combinatorial 
tool with several applications. As a result it appeared in 1975 
in a paper  by  Erd\H{o}s and Lov\'{a}sz \cite{erdos1975problems}.
Recently,  a lot of work has been focused on constructive proofs of LLL.
A major step in this direction has been the constructive proof of  Moser \cite{moser2009constructive}
and Moser and Tardos \cite{moser2010constructive}.
In this paper we give a constructive proof of the so called {\sl Asymmetric Lov\'asz Local Lemma}
that is based on the original approach  from~\cite{moser2009constructive} 
and the ideas from~\cite{KPT15}.

\vspace{-4mm}
\paragraph{The variable setting framework.}
We
 work in a framework known as the {\em variable setting}, which was used in~\cite{moser2009constructive}.
Let $X_i, i=1, \ldots,n$ be mutually independent random variables on a common probability space, taking values in the sets $D_i, i=1, \ldots, n$, respectively. 
Let also $E_j, j=1, \ldots,m $ be a sequence of events,  each depending on a sequence of the random variables $X_i$. We define the {\em scope} $e^j$ of event $E_j$ to be the minimal subset of variables such that one can determine whether $E_j$ is satisfied or not knowing only their values, i.e., event $E_j$ depends only on the values of the variables of $e^j$. ).
With every sequence of events ${\cal E} = E_1, \ldots E_m$ we associate a unique graph $G_{{\cal E}}$ called the {\em dependency graph} of ${\cal E}$ which is defined as follows: $V(G_{{\cal E}})=\{1,\ldots,m\}$ and  for every $i,j \in \{1, \ldots, m\}$, we have $\{i, j\}\in E(G_{{\cal E}})$ if and only if $e^i \cap e^j \neq \emptyset$.
For $j=1, \ldots, m$, we define the {\em neighborhood} of event $E_j$, denoted by $N_j$, to be the neighborhood of the vertex $j$ in the dependency graph, i.e., $N_j = \{ i \in \{1, \ldots, m\} \mid \{i, j\}\in E(G_{{\cal E}})\}$ (observe that $i \in N_i$)\medskip

\noindent {\bf Asymmetric Lov{\'a}sz Local Lemma.}
The main goal of this work is to present an alternative
 algorithmic proof of the following theorem:

\vspace{-2mm}
\begin{theorem}[Asymmetric Lov{\'a}sz Local Lemma]\label{thm:LLL}
If there exist $\chi_1,\chi_2,\ldots,\chi_m\in(0,1)$ such that $  \forall i \in \{1,\dots, m\}\ Pr(E_i)\leq \chi_i \prod_{j\in N_i} (1-\chi_j)$
then $Pr[\overline{E_1}\wedge \overline{E_2}\wedge \cdots \wedge\overline{E_m}]>0$, i.e., there exists an assignment to the variables $X_i$ for which none of the events $E_i$ hold.
\end{theorem}
The original proof of Theorem  \ref{thm:LLL}, first presented in this form in   \cite{Spencer197769}, was  non-constructive, but was given for arbitrary events, i.e. without the assumption that the events depended on independent random variables. Below, we will give  an algorithmic proof Theorem  \ref{thm:LLL} within the variable framework, where each event depends on a subset of the variables.

\begin{figure}[h]
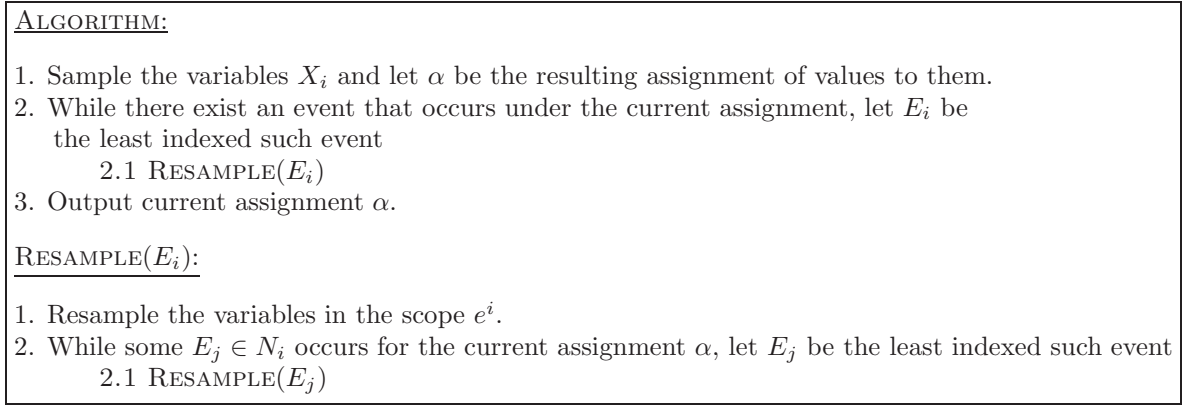
\begin{center}
\fbox{\small \begin{minipage}{4in}
\noindent \underline{\emph{\sc Algorithm}:}
\begin{program}
1. Sample the variables $X_i$ and let $\alpha$ be the resulting assignment  of values to them. \\

2. While there exist an event that occurs under the current assignment,  let $E_i$ be\\ \hspace*{0.4cm} the least indexed  such event \\
     \hspace*{1cm} {\sc 2.1 Resample}($E_i$) \\
     
3. Output current assignment $\alpha$.
\end{program}
\noindent \underline{\emph{\sc Resample}($E_i$):}
\begin{program}
1. Resample the variables in the scope $e^i$.\\
2. While some $E_j \in N_i $ occurs for the current assignment $\alpha$, let $E_j$ be the least indexed such event\\
 \hspace*{1cm} {\sc 2.1 Resample}($E_j$)
\end{program}
\end{minipage}}
\caption{{\bf \textit{Randomized sampling}}  algorithm}\label{fig:alg}
\end{center}
\label{this:fig}\end{figure}

We analyze the algorithm presented in Figure~\ref{fig:alg}, which can be derived directly from the one given by Moser in~\cite{moser2009constructive}, and which is the same that we used in~\cite{KPT15}. Our goal is to prove that the probability of {\emph {\sc Algorithm}} making many steps is sub-exponential to the number of steps.

We call {\em phase}, the execution period within a {\em root} call of {\sc Resample} (line 2.1 of {\sc Algorithm} in Figure~\ref{this:fig}) and we will refer to calls of {\sc Resample} from within another {\sc Resample} execution as {\em resample} calls. In these terms, our task is to bound the probability that the \emph{\sc Algorithm} makes at least $n$ \emph{\sc Resample} calls. To do this, we first show that the number of phases in any execution is bounded, which can be derived form the following lemma:




\vspace{-2mm}
\begin{lemma}\label{lem:progress}
Consider an arbitrary call of \emph{\sc Resample($E_i$)}. Let $\mathcal{E}$ be the set of events that do not occur at the beginning of this call. Then, if the call terminates, events in $\mathcal{E}$ will also not occur at the end of the call.
\end{lemma}


The above lemma implies that the number of phases in any execution is bounded as it states that the events that do not occur at the start of a {\em{\sc Resample}} call also do not occur after its end. Thus, we have the following corollary:

\vspace{-2mm}
\begin{corollary}\label{cor:phases}
There are at most $m$ phases in any execution of \emph{\sc Algorithm}.
\end{corollary}

\vspace{-2mm}
Let us now examine the probability distribution of the variables after a resampling caused by a call of \emph{\sc Resample($E_i$)}.

\vspace{-2mm}
\begin{lemma}[Randomness lemma]\label{lem:randomness}
Let $\alpha$ be a random assignment sampled from the probability distribution of the  variables $X_1, \ldots, X_n$ and $E_i$ an event. Let $\alpha'$ be the assignment obtained from $\alpha$ by resampling the variables in $e^i$ if $E_i$ occurs for $\alpha$, and let $\alpha'$ be $\alpha$ otherwise. Then, conditional that $E_i$ occurs under $\alpha$, the distribution of $\alpha'$ is the random distribution of assignments sampled from all variables i.e. it is the same as the distribution of $\alpha$. Therefore the probability that any event $E$ occurs under $\alpha'$ is equal to the probability that $E$ occurs under $\alpha$. 
\end{lemma}
\remove{
\begin{proof}
This immediately follows from the principle of deferred decisions. One only needs to consider the values of the variables in $e^i$ after the resampling and since these are sampled from the same distribution, $\alpha'$ can be seen as sampled from the same distribution as $\alpha$. Notice that without the conditional that $E_i$ occurs under $\alpha$, the lemma is not, in general,  correct as then the resampling does not necessarily take place.
\QED \end{proof}
}

\vspace{-2mm}
Now we provide two definitions identical to the ones given in~\cite{KPT15} which are analyzed there.

\vspace{-2mm}
\begin{definition}
\label{witness}
A sequence of events $\mathcal{E}_1, \ldots, \mathcal{E}_k$ is called a {\em witness sequence} if the   first $k$ {\sc Resample} calls (recursive or root)  of {\emph {\sc Algorithm}} are applied to $\mathcal{E}_1, \ldots, \mathcal{E}_k$, respectively.
\end{definition}

\noindent Now let $\hat{P}_n$ be the probability that {\sc Algorithm} performs at least $n$ {\sc Resample} calls. It follows that  $\hat{P}_n = \Pr\left[\text{ there is some witness sequence }  \text{of length } n \text{  }  \right].$
\smallskip

\vspace{-3mm}
\begin{definition}
\label{valid}
A sequence of events $\mathcal{E}_1, \ldots, \mathcal{E}_k$ is called a {\em valid sequence} if  

\noindent $\bullet$ there is a rooted  forest with at most $m$ trees labeled with the events in the sequence so that  the order of the events in the sequence coincides with  the preorder of  the labels of the forest (the same label may appear more than once),  \\
$\bullet$ the label of a non-root node $v$  in the forest is a neighbor of the label of the parent of $v$,\\ 
$\bullet$ the indices of the labels of the successive  children of any node are strictly increasing; the same is true for the indices of  the successive root labels of the forest, and   \\
$\bullet$ $\mathcal{E}_i$ occurs for the assignment $\alpha_i, i=1, \ldots, k$, 
where $\alpha_1$ is obtained by sampling $X_1, \ldots, X_n$ 
and  $\alpha_{i+1}, i = 1, \ldots, k-1$ is obtained by resampling the variables in $e^i$.  
 \end{definition}

\noindent If we now define: $P_n = \Pr\left[\text{ there is some valid sequence }  \text{of length } n \text{ }  \right]$
and assuming  $P_0=1$, we have clearly that  $\hat{P}_n \leq  P_n.$
With the purpose of bounding $P_n$, we define $Q_{n,i}$ to be the probability that there is some valid sequence starting from $E_i$ and of length $\geq n$.
Our target from now on is to upper bound the numbers $Q_{n,i}$.

%

\vspace{-4mm}
\paragraph{Recurrence relations.}
We will need the following lemma (proof is omitted):

\vspace{-2mm}
\begin{lemma} For every $n$ and every $i \in \{1, \ldots, m\}$, assuming that $N_i = \{i_1, \ldots i_l\}$, the numbers $Q_{n,i}$ satisfy the following equation:
\vspace{-2mm}

\begin{equation}
\label{rec}
Q_{n,i} = Pr(E_i) \Bigg( \hspace{0.2cm} \mathlarger{\mathlarger{\sum}}_{n_1+ \ldots +n_l = n-1} Q_{n_1, i_1}  Q_{n_2, i_2} \cdots Q_{n_l, i_l}\Bigg)
\end{equation}
\end{lemma}
%
%

\noindent Denote  $Q_i(z) = \sum_{n=1}^{\infty} Q_{n,i}\cdot z^n$.
\noindent Multiplying \eqref{rec} by $z^n$ we get 
\begin{equation}
\label{pregen}
Q_{n,i}z^n =z Pr(E_i) \Bigg( \hspace{0.2cm} \mathlarger{\mathlarger{\sum}}_{n_1+ \ldots +n_l = n-1} Q_{n_1, i_1}z^{n_1} \cdot Q_{n_2, i_2}z^{n_2} \cdots Q_{n_l, i_l}z^{n_l}\Bigg)
\end{equation}

\noindent and we add over all $n$ and to get $Q_i(z) = z\cdot  Pr(E_i) \prod_{j \in N_i} \big(Q_j(z) + 1\big)$.
\noindent Now we have obtained a system of equations ($\overline{Q}=(Q_1,Q_2,\ldots,Q_m)$):
$Q_i(z)=zf_i(\overline{Q})$
where $f_i(\overline{Q})=\chi_i  \prod_{j\in N_i} (1-\chi_j)(Q_j+1).$
We will use Theorem 2 from~\cite{KPT15}. To do this, we need to make a generalization. Instead of our functions in one variable, we will generalize in $m$ variables. This will grant us a unique solution in a more general system, and since we know that our system also does have a solution it must be the same given that $x_1=x_2=...=x_m$. So we consider the same system where in each equation $z$ is substituted by $t_i$ and $Q_i$'s are generating functions on $t=(t_1,...,t_m)$. 
Using directly the main result of Bender and Richmond in~\cite{BenderR98} (Theorem 2), what we need to show is that the following quantity (which will be $Q_s(\mathbf{t}$)) is bounded by $\rho^n$ where $\rho<1$ and $\sum_{i=1}^{m} n_i=n$, probably ignoring polynomial factors involving $n$ and $m$ since they are asymptotically irrelevant ($g$ is the projection function):
$ \frac{1}{\prod n_i} [ \mathbf{x}^{\mathbf{n-1}} ] \sum_{T} \frac{d(g,f_1^{n_1},\ldots,f_m^{n_m})}{dT}.$
\noindent where the sum is over all rooted trees with edges directed towards $0$.
We observe that since the $\frac{1}{\prod n_i}$ factor is irrelevant, and that also the number of trees is a function of $m$, we just need to bound from above the term of the sum for an arbitrary tree.
Let $T$ be a tree. We need to upper bound the following: $ [ \mathbf{x}^{\mathbf{n-1}} ] \prod_{j\in V(T)} \Bigg\{ \bigg( \prod_{(i,j)\in E(T)} \frac{d}{dx_i} \bigg) f_j(\mathbf{x}) \Bigg\}.$
Now, we substitute the values of $f_i$ and we group corresponding to $i$, meaning that we  group together the factors $\chi_i$, $1-\chi_i$, and $x_i+1$. First, observe that if in $T$ vertex $0$ has 
a child (by child of a vertex we will mean another vertex from which there exists an edge directed towards it) other than $s$, then the product would be zero (since the derivative of $g(\mathbf{x})=x_s$ would be zero) and the same holds if 
a vertex $i$ has a child not in its neighborhood. So we will restrict our attention to the case that $T$ satisfies these constraints.  \\
All $\chi$'s remain as they were since they are constants. Notice that without the existence of the derivatives, the product would be:
$ \prod_{i=1}^m \Bigg\{ \chi_i^{n_i}\cdot (1-\chi_i)^{\sum_{j \in N_i} n_j} \cdot (x_i+1)^{\sum_{j \in N_i}n_j} \Bigg\}.$
In the actual product, the exponent of $x_i+1$ will be reduced by $1$ for every $\frac{d}{dx_i}$. But $T$ is a tree, so every node $i$ has outdegree one (except from the $i$ of the projection function but it will become obvious later that is safe to assume that this exponent is also reduced by $1$). So every exponent is reduced exactly by $1$.
So, ignoring $n_k$'s (which is safe as they are bounded by $n^m$), we need to bound:
$ [ \mathbf{x}^{\mathbf{n-1}} ] \prod_{i=1}^m \Bigg\{ \chi_i^{n_i}\cdot (1-\chi_i)^{\sum_{j \in N_i} n_j} \cdot (x_i+1)^{\sum_{j \in N_i} n_j-1} \Bigg\} $
But this, by binomial theorem, is: $ \prod_{i=1}^m \Bigg\{ \chi_i^{n_i}\cdot (1-\chi_i)^{\sum_{j \in N_i} n_j} \cdot {\sum_{j \in N_i} n_j-1 \choose n_i-1} \Bigg\}.$
Obviously, by the identity ${n \choose k}=\frac{n}{k} {n-1 \choose k-1}$ and by ignoring again non-exponential factors, this is:
\begin{eqnarray*}
& = &\prod_{i=1}^m \Bigg\{ \chi_i^{n_i}\cdot (1-\chi_i)^{\sum_{j \in N_i} n_j} \cdot {\sum_{j \in N_i} n_j \choose n_i} \Bigg\}  \\
 & = & \prod_{i=1}^m (1-\chi_i)^{n_i} \cdot \chi_i^{n_i}\cdot (1-\chi_i)^{\sum_{j \in N_i} n_j - n_i} \cdot {\sum_{j \in N_i} n_j \choose n_i} \\
 & < & \prod_{i=1}^m (1-\chi_i)^{n_i}
\end{eqnarray*}
\vspace{-2mm}

\noindent The last inequality is derived by expanding $1=\big(\chi_i + (1-\chi_i)\big)^{\sum_{j \in N_i} n_j}$ and noticing that the term we bounded is just the $n_i^{th}$ term of the sum (which is a sum of nonnegative terms since $\chi_i \in (0,1)$, so any of its terms is less than the sum itself). Now, let $M=\max\{ (1-\chi_i) \}$ and observe that $M<1$, since $1-\chi_i<1$ for every $i$. Finally, we have:
$\prod_{i=1}^m (1-\chi_i)^{n_i} \leq  \prod_{i=1}^m M^{n_i} = M^{^{\sum n_i}} = M^n.$

To sum up, we proved that the probability of \emph{\sc Algorithm} making more than $n$ steps is sub-exponential in $n$, so we proved Theorem \ref{thm:LLL}.

\vspace{-3mm}
{\footnotesize
\bibliographystyle{plain}

}

\end{document}